\documentclass[twocolumn,aps,prl,showpacs,raggedbottom,nobalancelastpage,superscriptaddress]{revtex4-1}

\usepackage{graphicx}
\usepackage{amsmath}
\usepackage{amsfonts}

\begin{document}
\title{Class $\mathrm{D}$ spectral peak in Majorana quantum wires} 

\author{Dmitry Bagrets and Alexander Altland}

\affiliation{Institut f{\"u}r Theoretische Physik,
Universit{\"a}t zu K{\"o}ln, K{\"o}ln, 50937, Germany}

\begin{abstract}
  Proximity coupled spin-orbit quantum wires purportedly support
  midgap Majorana states at critical points. We show that in the
  presence of disorder  these systems generate a
  second bandcenter anomaly, which is of different physical origin but
  shares key characteristics with the Majorana state: it is narrow in
  width, insensitive to magnetic fields, carries unit spectral weight,
  and is rigidly tied to the band center. Depending on the parity of
  the number of subgap quasiparticle states, a Majorana mode does or
  does not coexist with the impurity peak. The strong
  'entanglement' between the two phenomena may hinder an unambiguous
  detection of the Majorana by spectroscopic techniques.
\end{abstract}
\maketitle
More than seven decades ago, the Majorana fermion was postulated as a
neutral variant of the Dirac particle~\cite{Majorana:1937qy}. While the Majorana has never
been observed in the canonical environments of particle physics,
recent proposals\cite{Kitaev:2001,Oreg:2010,Lutchyn:2010} suggest a realization as a boundary state
of one dimensional topological superconductors. The perspective to
realize, and potentially apply topologically protected Majorana
particles in condensed matter settings, has sparked a wave of theoretical~\cite{Alicea:2012,Flensberg:2010,Alicea:2011,Fidkowski:2011,Stoudenmire:2011,Lutchyn:2011,Sela:2011,Akhmerov:2011,
Brouwer:2012,Kells:2012,Heck:2012} and experimental~\cite{Mourik:2012, Rokhinson:2012, Deng:2012, Heiblum:2012}
activity. 

Recently, three experimental groups have reported signatures of Majorana 
states in semiconductor quantum wires. 
In these experiments, proximity coupled semiconductor quantum wire
purportedly hosting a Majorana particle are probed by tunneling
spectroscopy. Evidence for the presence of the particle is drawn from
the observation of a zero energy spectral peak. It has been
verified that the presence of the peak hinges on parametric conditions
necessary for the formation of a Majorana particle.

In this Letter we argue that the quasi one-dimensional quantum wires
presently under consideration are prone to the formation of a
\textit{second} zero energy peak, caused by strong midgap quantum
interference. The latter structure, here called (class $\mathrm{D}$)
spectral peak for brevity~\footnote{Strictly speaking, the system is
in class $\mathrm{D}$ or $\mathrm{B}$ depending on whether a
Majorana is absent or present. We stick to the designation
'$\mathrm{D}$' for simplicity.}, shares key universal signatures
with the Majorana peak, to the extent that disentangling the two
phenomena may be
difficult: it is (i) tied to  zero energy, (ii) 
affected by temperature, but not (iii) by magnetic field, (iv) relies
on parametric conditions similar to those required by the Majorana, 
(v) carries unit spectral weight (a single quasi
'state'), (vi) is independent of other midgap structures, such as
Kondo resonances or Andreev bound states, shows (vii) sensitivity to
the parity of channel number, however, unlike the Majorana,
(viii) relies on the presence of moderately weak disorder.

\begin{figure}[h]
  \centering
 \centerline{\includegraphics[width=8.5cm]{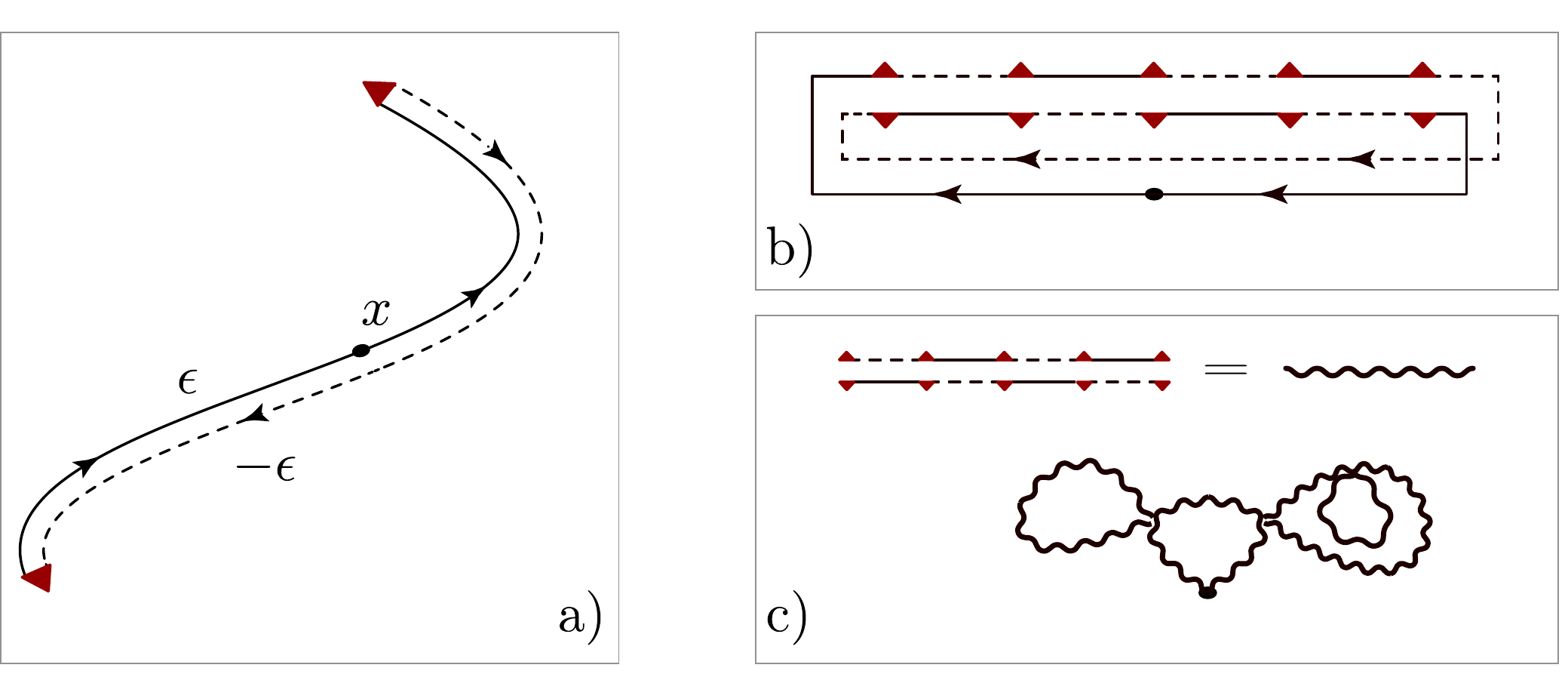}}
  \caption{a) Phase coherent scattering process $\mathrm{p}\to
    \mathrm{h}\to \mathrm{p}$, b) low energy diffusion mode formed by
    succession of such processes, c) cartoon of a higher order quantum
    interference process involving self-interacting diffusion modes.
  \label{fig:0}}
\end{figure}

\begin{figure}[h]
\centering
\centerline{\includegraphics[width=7.5cm]{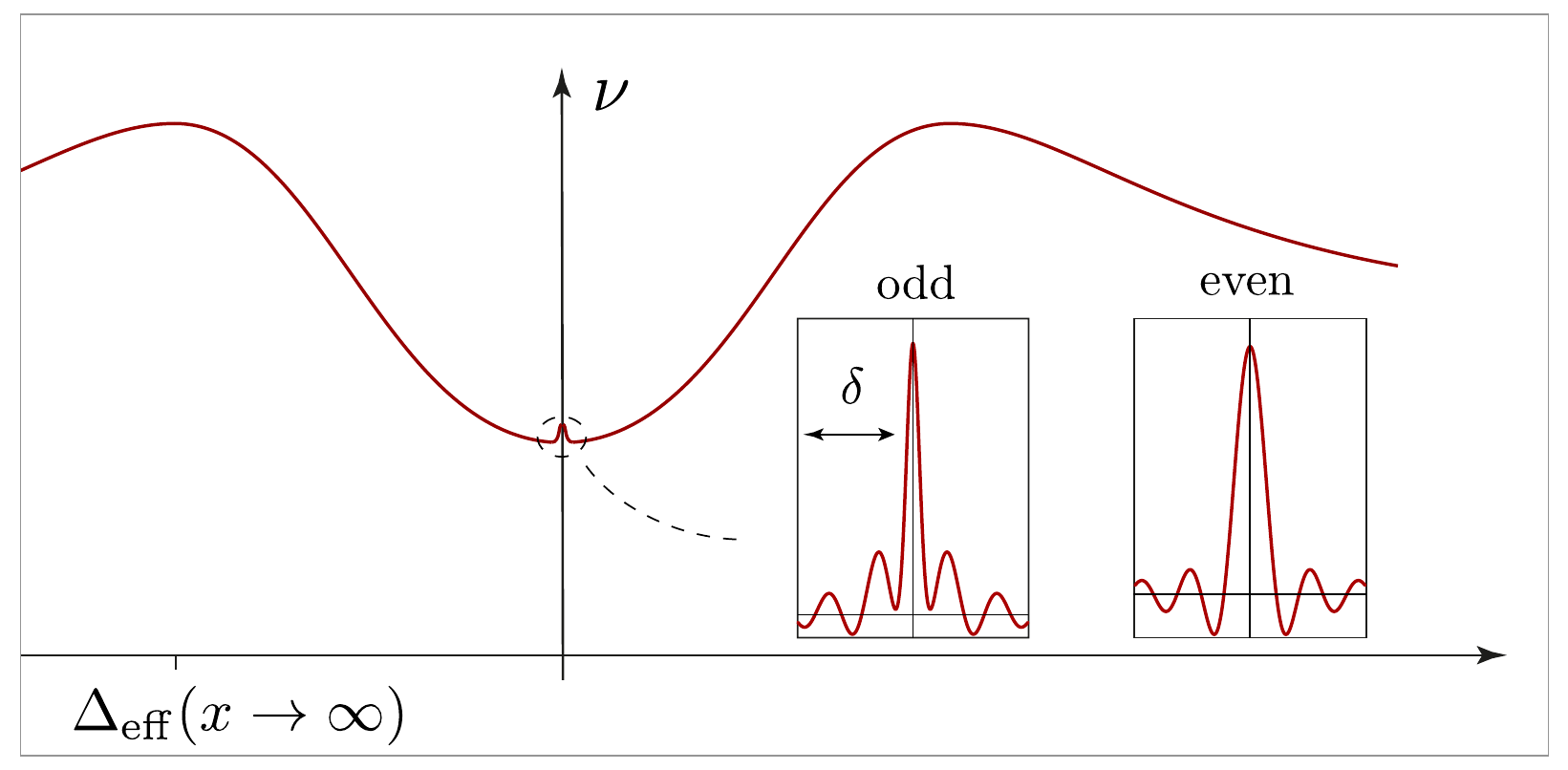}}
\caption{Schematic profile of spectral density. The profile of the band 
center anomaly depends on the parity of the number of channels,
$N$. Only for odd $N$ the system supports a genuine Majorana level which (here represented as
a broadened $\delta$-singularity.)
\label{fig:1}}
\end{figure}

\textit{Qualitative discussion and main results. ---} To start with,
let us qualitatively discuss the physics underlying the spectral
peak. The spin orbit and proximity coupled quantum wires presently
under discussion form a variant of a spinless superconductor, more
precisely a system of class $\mathrm{D}$~\cite{Altland:1997}.  At the
critical configuration separating a topologically trivial and
nontrivial state (which will be the habitat of the Majorana fermion if
parameters are tuned in a space-like manner) the superconductor goes
gapless. The presence of a Majorana particle signifies in a band
center anomaly of the quasiparticle density of states (DoS). Quantum
mechanically, the DoS at energy $\epsilon$ (relative to the system's
chemical potential, $\mu$) is given by $ \rho(\epsilon) =-{1\over
  \pi}\int dx\,\text{Im tr}(G^+(x,x;\epsilon))$, where the Green
function $G^+(x,y;\epsilon)\equiv\langle x| (\epsilon+i0+ \mu
\sigma_3^{\mathrm{ph}}-\hat H)^{-1}|y\rangle$, $\sigma_3^\mathrm{ph}$
is a Pauli matrix in the superconductor particle hole (\textrm{ph})
space, and \textrm{'tr'} is a trace over that space. Semiclassically,
 the matrix elements $G^{\mathrm{pp}}(x,x;\epsilon)$ (or
$G^{\mathrm{hh}}(x,x;\epsilon)$) may be interpreted as return
amplitudes of quasi-particles, in an environment governed by the
Bogoliubov-de Gennes Hamiltonian $\hat H$ of the system. Much like in
a gapless superconductor~\cite{Altland:2000}, the DoS in the
proximity quantum wire is affected by the scattering of low energy
excitations off spatial fluctuations of an order parameter
amplitude, $\Delta$. To understand the physics of this
mechanism~\cite{Altland:1997}, imagine a quasi-particle of energy
$\epsilon$ emanating from the point $x$. The particle has
the option to scatter off the order parameter $\Delta$ into
a quasi-hole of energy $-\epsilon$, which by a second scattering
process may get converted back to a particle returning to
the point of origin. For low excitation energies, the particle and
hole stretches involved in this process interfere constructively to
give a robust (scattering phase insensitive) contribution to the
return amplitude. Summation over repeated $\mathrm{ph}$ conversion
processes generates an effective diffusion mode~\cite{Altland:1997}
similar to the diffuson and Cooperon modes of conventional disordered
metals. The increase of the effective quasi-particle gap away from the
critical point confines the support of this mode to a small region in
space -- a disordered gapless superconductor 'quantum dot' -- where it
scales as $\sim \delta/(2\pi\epsilon)\equiv s^{-1}$, $\delta$ being
the dot's effective level spacing. In disordered electronic systems,
singularities of this type are generally cut off by the nonlinear
'self-interaction' of diffusion modes (weak localization turning into
strong localization, or level repulsion being prominent examples of
such nonlinearities). In the present context, similar nonlinearities
lead us to the result
\begin{align}
\label{eq:3}
  N\;\mathrm{even}:\quad&{\rho(s)\over \rho_0}=1+{\sin(s)\over s}, \qquad
  s=2\pi
  {\epsilon\over \delta}\cr
  N\;\mathrm{odd}:\quad &{\rho(s)\over \rho_0}=1-{\sin(s)\over s}+\delta\left({s\over 2\pi}\right),
\end{align}
where $N$ is the number of channels in the system, $\delta$ the
average quasiparticle level spacing, $\rho_0\equiv \delta^{-1}$, and
the $\delta$-function is the contribution of the Majorana
state. Multiple quantum interference multiplies the single diffusion
mode by a phase factor $\pm \sin(s)$, where the overall sign depends
on the channel number parity.  The ensuing profiles of the band center
DoS are shown in the inset to Fig.~\ref{fig:1}. Notice the
superficially similar structures for even and odd channel numbers,
while only in the latter case a genuine Majorana mode is present. For
even channel numbers, a narrow width ($\sim \delta$) peak is solely
formed by class $\mathrm{D}$ diffusion modes. In either case, the
anomalous contribution to the DoS integrates to the spectral weight
$\int d\epsilon \, (\rho(\epsilon)-\rho_0) = {1\over 2}$. This means
that tunneling spectroscopy limited in resolution to scales larger
than $\delta$ may not be able to unambiguously identify the Majorana
state in situations where quantum interference is effective.  An
external loss of coherence, e.g. due to quasiparticle interactions,
will suppress the class $\mathrm{D}$ peak, although a power law
correction $\sim |\epsilon|^{-1}$ to the density of states
$\rho(\epsilon)$ above an effective cutoff energy should remain
visible. Finally notice how the phenomenon hinges on the presence of
disorder: impurities render the low energy dynamics diffusive, and
they provide the basis for the formation of an 'impurity band' of
quasiparticles centered around the gap closing point in energy and
space. However, we will argue below that even weak scattering
$\tau^{-1}\gtrsim \delta$ suffices to generate the effect. The rest of
the paper is devoted to a derivation of Eq.~\eqref{eq:3}. We will also
discuss physical bounds on 
the spatial extension of the impurity quantum dot, the relevant
disorder strengths, and other system parameters.

\textit{Model. ---} Majorana states can be realized in different
variants of class $\mathrm{D}$ topological superconductors. For
definiteness, we here consider the case of a proximity coupled helical
liquid~\cite{Sela:2011} subject to a smoothly varying magnetic
field. However, we will argue below that our main conclusions are not
rigidly tied to this setup. A helical liquid is formed by a system of
left- and right-moving fermions, $\psi_{L,\uparrow}\equiv \psi_L$ and
$\psi_{R,\downarrow}\equiv \psi_R$ carrying spin up and down,
respectively. (In the semiconductor setting, these modes form at the
intersection of fermion bands shifted in momentum by strong spin-orbit
interaction.) The coupling to a wire axis Zeeman field, $B$, and a
proximity $s$-wave order parameter then generates the effective
Hamiltonian $\hat H_0=\sum_{a=1}^N\int dx({\cal H}^a_K+{\cal
  H}^a_B+{\cal H}^a_\Delta)$, where ${\cal H}^a_K=\sum_{C=L,R} \,\bar
\psi^a_C is_C v_F \partial_x\psi^a_C$, ${\cal H}^a_B= \,B\bar \psi^a_L
\psi^a_R + \mathrm{h.c.}$, ${\cal H}^a_\Delta= \,\Delta \psi^a_L
\psi^a_R + \mathrm{h.c.}$, $v_F$ is the Fermi energy, and $s_C=(+/-)1$
for $C=L/R$. We have also introduced an index $a=1,\dots, N$, which
accounts for the option of multiple bands $\psi_C^a$ crossing the
Fermi energy. Differences in the Fermi velocities of the bands are
absorbed in a rescaling of the fields. Ignoring the effects of band
coupling by spin orbit interaction, the effects of static disorder and
interaction, etc. $\hat H_0$ has the status of a null theory
describing the formation of a critical regime at $B \simeq \Delta$:
introducing new state vectors ('Majorana basis'), 
$ \eta^0\equiv
{1\over \sqrt 2}\left(
      \begin{smallmatrix}
 -\psi_R - \bar
  \psi_R\cr
        i\psi_L -i \bar
  \psi_L       
      \end{smallmatrix}
    \right),\eta^1\equiv 
{1\over \sqrt 2}\left(
      \begin{smallmatrix}
        i\psi_R -i \bar
  \psi_R\cr
 \psi_L + \bar
  \psi_L
      \end{smallmatrix}
    \right),
 $
and assuming reality of the order parameter, $\Delta\simeq B$, the Hamiltonian $\hat H_0$ assumes the form
  \begin{align*}
    \hat H_0=\sum_{\mu=0,1}\int dx\, \eta^{\mu T}
    \left(-iv_F \partial_x \sigma_3 +(B-(-)^\mu\Delta)\sigma_2
    \right)\eta^\mu,
  \end{align*}
where a summation over channel indices is implicit and the Pauli
matrices act in $LR$-space.  Notice that the matrix structure
sandwiched by the $\eta$'s is antisymmetric, which is the defining
condition of a class $\mathrm{D}$ (no symmetries other than
particle-hole symmetry) superconductor. The above representation
makes the formation of a low- $(\eta^0)$ and a high-energy sector
$(\eta^1)$ manifest. In the absence of perturbations, $\eta^{0,a}$
represent Majorana modes which go massless at the critical point
$\Delta_\mathrm{eff}\equiv \Delta-B=0$. Throughout, we will assume a
hierarchy of energy scales $M\equiv B+\Delta \gg
\Delta_\mathrm{eff}$, and on this basis we now turn to the
discussion of perturbations, $\delta \hat H$, to $\hat H_0$. Scattering
between the low- and the high-energy sector -- generated by
fluctuations of the chemical potential, or imaginary contributions
to the order parameter -- affects the low energy theory by
contributions of ${\cal O}(\delta H^2/M)$) and will be neglected
throughout. By contrast, real fluctuations of the order parameter,
or channel dependent differences in the field strength (indirectly
generated by inter-channel spin-orbit coupling) couple within the
low energy sector and have to be kept. We model such fluctuations in
terms of a phenomenological ansatz
$\Delta_\mathrm{eff}=\Delta_0(x)+W+V$, where the first term
describes a ramping of system parameters through a critical point at
$x=0$, where $\Delta_0(0)=0$ (cf. Fig. \ref{fig:1}), $W=\{W^{ab}\}$ is a matrix containing
deterministic contributions to the inter-channel coupling, and
$V=\{V^{ab}(x)\}$ represents static disorder, here assumed to be
Gaussian distributed, $\langle V^{ab}(x) V^{a'b'}(x')\rangle =
\delta^{aa'}\delta^{bb'}\delta(x-x') \gamma^2$. Throughout we will
limit our discussion to channels for which the low energy physics
around $x=0$ is disorder dominated, i.e. for which the
random potential $V>W$ masks the deterministic contribution. (If
this condition cannot be satisfied by a multiplet of channels, we
are down to $N=1$.\footnote{$N=1$ is a special case. For
abruptly varying control parameters (an 'interface'), the system then
supports only one level below the bulk gap, and our analysis does
not apply. For smoothly varying parameters
Anderson localization on length scales $l\sim \tau v_F$ may
become an issue. Although the detailed analysis is
beyond the scope of this paper, a band center peak should be
observable also in this case.}) 

\textit{$\sigma$-model. ---} Our focus throughout will be on regions
close to $x=0$ where the gap amplitude $|\Delta(x)|<V$ is small enough
for a class $\mathrm{D}$ impurity quantum dot to form.  Such
structures universally exhibit a zero energy spectral peak which we
aim to explore quantitatively. To this end, we consider the
supersymmetric functional integral $\int D(\bar \xi,\xi) \exp(i\int dx
\,\bar\xi(\epsilon^+-\hat H)\xi)$, where $ \hat H=-iv_F \partial_x
\sigma_3 - \Delta_\mathrm{eff}(x)\sigma_2$ is the low energy ($\mu=0$)
Hamiltonian, and $\xi=\{\xi_{C,\alpha}\}$
and $\bar \xi=\{\bar\xi_{C,\alpha}\}$ are integration variables with
complex ($\alpha=\mathrm{b}$) and Grassmann valued
($\alpha=\mathrm{f}$) components. 
The DoS can be obtained from the above
functional by differentiation w.r.t. suitably introduced sources,
which we make not explicit for notational simplicity.

\begin{figure}[h]
 \centering
 \centerline{\includegraphics[width=8.5cm]{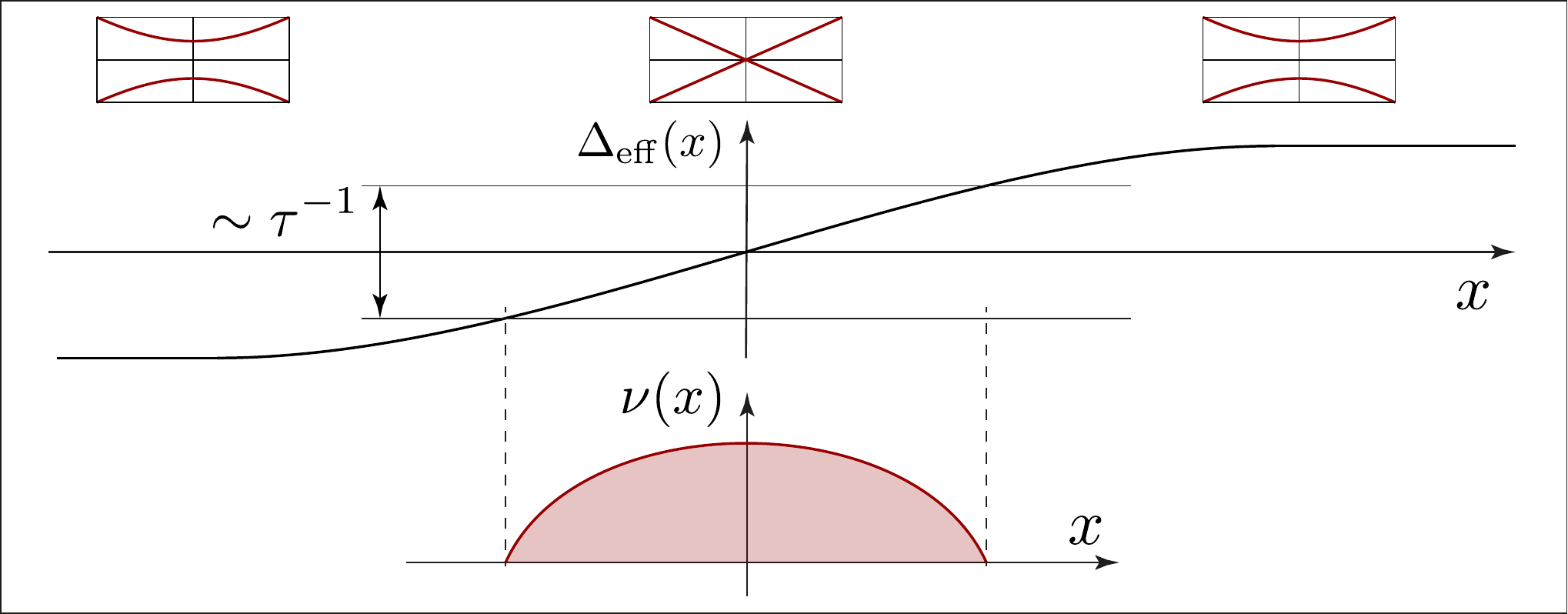}}
  \caption{Top: a sweep of an effective gap
    parameter generates a critical point at which the clean system goes
    gapless (cf. schematic dispersion relations shown in the
    insets.) Bottom: in the presence of disorder an impurity band of
    semicircular local average DoS forms.
  \label{fig:2}}
\end{figure}

We next average the generating functional over $V$, decouple the
ensuing term quartic in $\xi$ by Hubbard-Stratonovich transformation,
and integrate over $\xi$, to obtain the action $ S[X,\bar X]={1\over
  2\gamma^2} \int dx \, \mathrm{str}(X\bar X)+{1\over 2}
\mathrm{str\,ln}\big(\epsilon^+\tau_3 +X_S+X_A\sigma_3 -\hat
H_0\big)$, where 'str' is the supertrace~\cite{Efetov:1997} and $\hat
H_0=\hat H\big|_{V=0}$. Here we have introduced a new 2-component
'\textrm{cc}' space whose sectors $\sim (\psi,\psi^T)$ contains
states, $\psi$, and their Majorana-basis charge conjugates, $\psi^T$,
resp. The Pauli-matrix, $\tau_3$ acting in that space reflects the
different transformation behavior of $\hat H$ and $\epsilon^+$,
$(\epsilon^+-\hat H)^T = -(-\epsilon^+-\hat H)$.  The four component
supermatrices $X=\{X_{t \alpha,t'\alpha'}\}$, and $\bar X=X^\dagger
\sigma_3^\mathrm{bf}$ are Hubbard-Stratonovich fields, where the Pauli
matrix in \textrm{bf} space ensures convergence, $t,t'$ are indices in
$\mathrm{cc}$-space, and $X_{S,A}={1\over 2}(X\pm \bar X)$.
Following standard strategies~\cite{Efetov:1997}, we process the above
action by a variational approach. Employing the polar ansatz, $X=\bar
X=i \lambda T \Lambda T^{-1}$, where $T$ are rotation matrices to be
discussed momentarily and $\Lambda$ is a diagonal matrix carrying real
matrix elements $\pm 1$, we find that the modulus
$\lambda=\lambda(x)\ge 0$ is fixed by the variational equation $
\lambda(x) = \gamma^2 \mathrm{Im}\langle x| (i\lambda-\hat H_0)^{-1}|
x\rangle$, which, upon substitution of $\hat H_0$ becomes
\begin{align*}
  &\sum_a \int {dp\over 2\pi}\, {\gamma^2\over (v_F p)^2
    +\Delta^{a2}+\lambda^2}=\sum_a {\gamma^2\over
    v_F\sqrt{\Delta^{a2}+\lambda^2}}=1.
\end{align*}
Here we have assumed the matrix $\Delta_\mathrm{eff}\big|_{V=0}\equiv
\{\Delta^a\delta^{ab}\}$ diagonalized, and varying slowly enough in
space for the functions $\lambda(x)$ to adiabatically follow.
Inspection of the equation shows that solutions exist if
$|\Delta^a|<N \gamma^2/v_F$ on average over $a$. In the limiting case
of negligible channel dependence, $\Delta^a \simeq \Delta_0$, we
readily obtain
\begin{align*}
  \lambda(x)=\left(\left({1\over 2\tau} \right)^{2} - \Delta_0^2(x)\right)^{1/2},
\end{align*}
where $\tau^{-1} \equiv {2N \gamma^2/v_F}$ is the golden rule
scattering rate of the unperturbed spin-orbit quantum wire. The
interpretation of $G_0^+(\epsilon)\equiv (\epsilon^+ - \hat H_0 + i
\lambda)^{-1}$ as an averaged Green function then yields $\nu=2\tau
\nu_0 \lambda$ for the (position dependent) DoS per unit length at saddle point level,
where $\nu_0=N/(\pi v_F)$. From this quantity, we generate the average
DoS as $\rho_0=\int_{\nu(x)>0} dx\,\nu(x)$, where the integral is over the
region of support of the local DoS, 
$|\Delta_0(x)|<1/2\tau$ (cf. Fig. \ref{fig:2}.) In passing we note that for large energies
$|\epsilon|\gg \tau^{-1}$, the density of states assumes a form
reminiscent of a smeared BCS profile to be discussed in more detail
elsewhere~\footnote{D. Bagrets and A. Altland to be published.}.

Turning to the saddle point matrix $\Lambda$, we note that, as with
the standard $\sigma$-model~\cite{Efetov:1997}, the eigenvalue
structure of the bosonic sector $\Lambda_\mathrm{bb}= \tau_3$ is fixed
by the pole signature of the Green function. However, in the fermionic
sector we have a choice, $\Lambda_\mathrm{ff}=\pm \tau_3$, an
ambiguity intimately related to the formation of a Majorana
peak~\cite{Ivanov:2002}: in the limit of low energies, $\epsilon\to
0$, spatially uniform rotations $\Lambda\to T \Lambda T^{-1}\equiv Q$
leave the action invariant. The symmetries of our class $\mathrm{D}$
theory require~\cite{Zirnbauer:1996} $T_\mathrm{bb}\in
\mathrm{Sp}(2)/\mathrm{U}(1)$, while $T_\mathrm{ff}\in
\mathrm{O}(2)/\mathrm{U}(1)\simeq \Bbb{Z}_2$. This latter identity
means that in the fermionic sector we have only two \textit{discrete}
symmetry transformations, $T_\mathrm{ff}=\openone$, and
$T_\mathrm{ff}=\tau_1$, where the latter acts by an exchange
$T_\mathrm{ff}\tau_3 T^{-1}_\mathrm{ff}=-\tau_3$ of the diagonal
saddle points. There are no continuous transformations $T$
interpolating between $\Lambda_1$ and $\Lambda_2$, which means that
the saddle point manifold, ${\cal M}\equiv {\cal M}_1\cup {\cal M}_2$
splits into two disjoint components, where ${\cal M}_s, s=1,2$ is
generated by action of all transformations $T$ with
$T_\mathrm{ff}=\openone$ on the saddle point $\Lambda=
\Lambda_1\equiv\tau_3$ and $\Lambda= \Lambda_2\equiv\tau_3 \otimes
\sigma_3^\mathrm{bf}$, resp.

We finally substitute slowly fluctuating 
configurations $X \to i \lambda Q$ into the action and
expand in small symmetry breaking contributions $\sim \epsilon^+Q $
and $\partial Q$. The expansion follows standard
procedures~\cite{Efetov:1997, Altland:2000} and generates the
effective action
\begin{align*}
  S[Q]=\int dx\, {\pi\nu(x)\over 8} \mathrm{tr}\left(D(x)\partial
    Q\partial Q+4\epsilon \tau_3 Q\right)+S_\mathrm{par}[Q],
\end{align*}
where $S_\mathrm{par}[Q]={1\over 2}\mathrm{str\, ln}(\Lambda-\hat
H_0$) is the Pfaffian of the model Hamiltonian at saddle point level,
and $D = { v_F\over \pi \nu}\sum_{a}{\lambda^2\over (\lambda^2 +
  \Delta^{a2})^{3/2}}$ plays the role of an effective diffusion
coefficient, which in the isotropic limit $\Delta^a\simeq \Delta_0$
simplifies to $D\sim (v_F \tau)^2 \lambda$. Notice how in the disorder
dominates regions $|\Delta| \ll \lambda$, $D\sim v_F^2\tau$ reduces to
the diffusion constant of a metallic system, while $D\to 0$ at the
boundaries $\lambda\to 0$ of the metallic domain. The structure of the
action above indicates that fluctuations of $Q$ generate the diffusive
soft modes discussed in connection with Fig. \ref{fig:0}.

\textit{Quasiparticle 'quantum dot'. ---} For energies
$\epsilon\lesssim \langle D\rangle /L^2$ lower than the minimum
excitation energy of spatially varying $Q$-fluctuations
($L$ is the extension of the metallic domain $\lambda>|\Delta_0|$, and
$\langle D\rangle$ is the typical value of the diffusion coefficient),
the action is dominated by homogeneous modes $Q=\mathrm{const}$. In
this regime, the system truly behaves like a
'quantum dot', and the action collapses to
\begin{align}
\label{eq:1}
  S[Q]={\pi  \epsilon^+\over 2\delta} \mathrm{tr}(Q \tau_3)+S_\mathrm{par}[Q].
\end{align}
Here, $S_\mathrm{par}$ is a topological term~\cite{Ivanov:2002} 
which in the present context assumes the form
$S_\mathrm{par}[Q]=S_\mathrm{par}[ \Lambda_1]=0$ and
$S_\mathrm{par}[Q]=S_\mathrm{par}[
\Lambda_2]=\,\mathrm{tr}\,\mathrm{ln}(G_0^+(0)-G_0^-(0))$ for $Q\in
\mathcal{M}_{1,2}$ resp. Using the particle-hole symmetry of the
spectrum, the latter expression evaluates to 
$S_\mathrm{par}[\Lambda_2]=-i\pi{\cal N}_\mathrm{tot}$ where $N_\mathrm{tot}=\sum_a N_a$, 
and  $N_a$ is the total number of states carried by mode $a$. While this number
 does not carry physical meaning by itself, $\exp(-S_\mathrm{par})=(-)^{N_\mathrm{tot}}$ responds
only to the \textit{parity} of $N_\mathrm{tot}$. Now, robust arguments
show that $N_a=(2k+1)$ is generally odd~\footnote{The point is that each mode, $a$, is
    described by an Hamilton operator obeying the symmetry $[\hat
    H,\sigma _1]_+=0$. This chirality implies that all eigenstates of
    non-vanishing energy come in degenerate pairs. For gap functions
    $\Delta _\protect \mathrm {eff}$ possessing a critical point,
    $\Delta _\protect \mathrm {eff}(0)=0$, the channel supports a
    non-degenerate zero energy state, i.e. the total number of states,
    $N_a$, is odd.}. 
This means that $(-)^{N_\mathrm{tot}}=(-)^N$ is susceptible
to the parity of the number of channels. The DoS resulting from Eq. (\ref{eq:1}), 
computed in Ref.~\cite{Ivanov:2002} by non-perturbative integration
over ${\cal M}_{1,2}$, is given by Eq. \eqref{eq:3}. 

While Eq. \eqref{eq:3} has been derived for a specific model, the high
degree of universality of the result --- dependence on no parameters
other than the ratio $\epsilon/\delta$ --- testifies to wider
applicability. Generally speaking, we expect the result to hold
provided $\delta\lesssim\tau^{-1}\ll \Delta$, i.e. the randomness
hybridizing a number of levels of the clean system around the critical point
$x\simeq 0$. This expectation is corroborated by the observation of
similar spectral profiles for phenomenological random matrix
models~\cite{Altland:1997}. (A straightforward numerical calculation
indeed shows that peaks similar to (\ref{eq:3}) form already for \textit{two}
randomly coupled levels.) Finally, random spectra are notorious for
their 'level rigidity', i.e. the weakness of sample-to-sample
fluctuations. Signatures of the anomaly in the 
\textit{ensemble averaged global DoS} are therefore expected to 
show in both the sample specific global and local DoS.  

Summarizing, we have shown that
in the presence of even weak disorder, a Majorana peak in class
$\mathrm{D}$ quantum wires will coexist with a spectral anomaly. The
two phenomena share key universal features, are of comparable
magnitude, and integrate to a spectral weight insensitive to the
parity of the channel number. All this makes us speculate that
disorder may be detrimental to an unambiguous observation of the
Majorana particle by spectroscopic means.

We thank F. von Oppen for discussions. Work supported by SFB/TR12 of the Deutsche Forschungsgemeinschaft.
 
\bibliography{my_biblio}
\bibliographystyle{apsrev4-1}

\end{document}